\documentclass[twocolumn]{aastex631}

\usepackage{lineno}
\usepackage{natbib}

\newcommand\flux{erg\,cm$^{-2}$\,s$^{-1}$}
\newcommand\lumi{erg\,s$^{-1}$}
\newcommand\nh{cm$^{-2}$}
\newcommand\first{4FGL J1910.7$-$5320}
\newcommand\second{4FGL J2029.5$-$4237}

\shorttitle{Multi-wavelength observation of 4FGL J1910.7$-$5320}
\shortauthors{Au et al.}

\graphicspath{{./}{figures/}}

\begin{document}

\title{MULTI-WAVELENGTH OBSERVATIONS OF A NEW REDBACK MILLISECOND PULSAR 4FGL J1910.7$-$5320}

\author{Ka-Yui Au}
\affiliation{Department of Physics, National Cheng Kung University \\
No. 1 University Road, Tainan City 70101, TAIWAN}

\author{Jay Strader}
\affiliation{Department of Physics and Astronomy, Michigan State University, East Lansing, MI 48824, USA}

\author{Samuel J. Swihart}
\affiliation{National Research Council Research Associate, National Academy of Sciences, Washington, DC 20001, USA \\ resident at Naval Research Laboratory, Washington, DC 20375, USA}

\author{Lupin C. C. Lin}
\affiliation{Department of Physics, National Cheng Kung University \\
No. 1 University Road, Tainan City 70101, TAIWAN}

\author{Albert K. H. Kong}
\affiliation{Institute of Astronomy, National Tsing Hua University \\
Hsinchu 30013, Taiwan }

\author{Jumpei Takata}
\affiliation{School of Physics, Huazhong University of Science and Technology \\ Wuhan 430074, People's Republic of China}

\author{Chung-Yue Hui}
\affiliation{Department of Astronomy and Space Science, Chungnam National University \\ Daejeon 34134, Republic of Korea}

\author{Teresa Panurach}
\affiliation{Department of Physics and Astronomy, Michigan State University, East Lansing, MI 48824, USA}

\author{Isabella Molina}
\affiliation{Department of Physics and Astronomy, Michigan State University, East Lansing, MI 48824, USA}

\author{Elias Aydi}
\affiliation{Department of Physics and Astronomy, Michigan State University, East Lansing, MI 48824, USA}

\author{Kirill Sokolovsky}
\affiliation{Department of Physics and Astronomy, Michigan State University, East Lansing, MI 48824, USA}

\author{Kwan-Lok Li}
\affiliation{Department of Physics, National Cheng Kung University \\
No. 1 University Road, Tainan City 70101, TAIWAN}

\correspondingauthor{Ka-Yui Au, Kwan-Lok Li}
\email{kyau@phys.ncku.edu.tw, lilirayhk@phys.ncku.edu.tw}

\begin{abstract}

We present the study of multi-wavelength observations of an unidentified \textit{Fermi} Large Area Telescope (LAT) source, \first, a new candidate redback millisecond pulsar binary. In the 4FGL 95\% error region of \first, we find a possible binary with a 8.36-hr orbital period from the Catalina Real-Time Transient Survey (CRTS), confirmed by optical spectroscopy using the SOAR telescope. This optical source was recently independently discovered as a redback pulsar by the TRAPUM project, confirming our prediction. We fit the optical spectral energy distributions of \first\ with a blackbody model, inferring a maximum distance of 4.1 kpc by assuming that the companion fills its Roche-lobe with a radius of $R = 0.7\,R_\sun$. Using a 12.6 ks \textit{Chandra} X-ray observation, we identified an X-ray counterpart for \first, with a spectrum that can be described by an absorbed power-law with a photon index of $1.0\pm0.4$. The spectrally hard X-ray emission shows tentative evidence for orbital variability. Using more than 12 years of \textit{Fermi}-LAT data, we refined the position of the $\gamma$-ray source, and the optical candidate still lies within the 68\% positional error circle. In addition to \first, we find a variable optical source with a periodic signal of 4.28-hr inside the 4FGL catalog 95\% error region of another unidentified Fermi source, \second. However, the $\gamma$-ray source does not have a significant X-ray counterpart in a 11.7 ks \textit{Chandra} observation, with a 3-$\sigma$ flux upper limit of $2.4\times10^{-14}$ \flux\ (0.3--7 keV). Moreover, the optical source is outside our updated \textit{Fermi}-LAT 95\% error circle. These observational facts all suggest that this new redback millisecond pulsar powers the gamma-ray source \first\ while \second\ is unlikely the $\gamma$-ray counterpart to the 4.28-hr variable.

\end{abstract}

\keywords{Millisecond pulsars (1062), Gamma-ray sources (633), Compact binary stars (283)}

\section{Introduction} \label{sec:intro}

Millisecond pulsars (MSPs) are neutron stars with a very short spin period on the order of one thousandth of a second. One widely accepted explanation, known as the recycling scenario, is that MSPs were in binaries with donor stars and the accretion from a companion in a binary system continuously transfers the angular momentum to the pulsar \citep[][and some MSPs still remain in binaries after the recycling process]{alpar}. If the binaries started the recycling phase at long orbital periods, then the MSPs will be formed with a complex evolution dynamics. In these cases, the donors are far away from the pulsars, it did not fill the Roche-lobes during the process until it became a (sub)giant, which typically leads to an MSP-He White Dwarf (WD) system \citep{tau,mspwd}. For the low-mass X-ray binaries (LMXBs) that started the recycling process in tighter orbits, special subclasses of pulsar binaries can be formed: redback and black widow MSPs. 

The two classes have compact orbits, and hence, short periods ($\leq$ 1 day) with very low-mass companions ($M_C\gtrsim$ 0.1 $M_\sun$ for redbacks and $<$ 0.1 $M_\sun$ for black widows; \citealt{roberts2011new,chen}). They are called redbacks and black widows in analogy to the conduct of real spiders: the female redback and black widow spider cannibalizes their companion after copulation, just like the central neutron stars create energetic pulsar winds and high-power radiation fields, which ablate the nearby companions. This scenario could explain how isolated MSPs are formed \citep{van}. In recent years, some redback MSPs, PSR J1227-4853 \citep{roy}, PSR J1023+0038 \citep{archi,pat,sta}, and M28I \citep{pap}, showed a transition between the LMXB state and the radio pulsar state, providing strong evidence supporting the recycling scenario of the MSPs formation \citep{alpar}.

Before the launch of the \textit{Fermi} Large Area Telescope (LAT), discovering redback and black widow MSPs was challenging because the material blown off from the companion by the high-power pulsar wind/radiation can hide the radio emission of the pulsar. Therefore, it is difficult to find the MSP radio pulsations in blind all-sky radio surveys. However, the GeV $\gamma$-ray emission is not affected by the obscuring material, making LAT a great tool to discover redback and black widow MSP candidates with follow-up observations in other wavelengths, given that many redback/black widow MSPs have $\gamma$-ray emission \citep{fermipulsar,swihart}. Nowadays, it is also possible to find MSP binaries by using optical data from all-sky surveys (ZTF and \textit{Gaia}; \citealt{gaia,ztf,62}).

LAT, the main instrument on the \textit{Fermi Gamma-ray Space Telescope}, is an imaging telescope for high-energy $\gamma$-rays in the range of 100 MeV--300 GeV \citep{fermi}. Using 12 years of LAT data, the \textit{Fermi}-LAT science team made the \textit{Fermi}-LAT Fourth Source Catalog Data Release 3 (4FGL-DR3; \citealt{fermi4-3}) to list the known properties of all the detected point-like and extended $\gamma$-ray sources of \textit{Fermi}-LAT. 4FGL (\citealt{fermi4}; the first data release) and 4FGL-DR3 share the same data analysis method and model for the Galactic interstellar emission. One of the major differences between them is that 4FGL-DR3 used four more years than 4FGL, and so the inferred parameters of the 4FGL-DR3 $\gamma$-ray sources are generally more accurate.

In this paper, two pulsar-like unassociated \textit{Fermi} sources, \first\ and \second, were investigated. For \first, we present the multi-wavelength observations, including the data from \textit{Fermi}-LAT, \textit{Chandra X-ray observatory}, the Southern Astrophysical Research (SOAR) telescope, and the Catalina survey, strongly suggesting that it is a redback MSP candidate. In addition, the Transients and Pulsars with MeerKAT (TRAPUM) team has independently found the radio pulsations associated with \first, which is in line with our result. We also present the $\gamma$-ray, X-ray, and optical analyses of \second. However, we find no evidence that this second unassociated GeV source is a pulsar binary system.

\section{Searching for spider MSP candidates with 4FGL-DR3/CRTS} \label{sec:searching}

Since the first catalog release of \textit{Fermi}-LAT, there have been numerous attempts to expand the redback and black widow MSP population by searching the LAT catalog for new pulsar systems (e.g., \citealt{j1653,searching,li2016,j0427}). These searches are based on the well-known fact that most MSPs have (i) stable $\gamma$-ray light curves on a monthly time-scale, and (ii) curved $\gamma$-ray spectra rather than a simple power-law. In this project,  we did not use the spectral and timing properties of the $\gamma$-ray sources for picking up candidates. We cross-checked the 4FGL-DR3 and the Catalina Surveys Southern periodic variable star catalogs of the Catalina Real-Time Transient Surveys \citep[CRTS;][]{crts,2017cata,fermi4-3} and selected possible $\gamma$-ray emitting compact binaries with the following criteria.

\begin{itemize}
 \item The 4FGL-DR3 $\gamma$-ray sources must be ``unassociated sources'' with detection significances higher than $5\sigma$ and high Galactic latitudes (i.e., $|b| > 5\degr$).
 \item For all the selected 4FGL-DR3 sources, the semi-major axes of the 95\% error circles must be smaller than $0\fdg1$ to minimize the contamination (i.e. unrelated CRTS sources).
 \item The periods of the selected CRTS variables must be shorter than 24 hours.
 \item The selected CRTS sources must be fainter than 15 mag. Redback and black widow systems are generally fainter than the threshold, although there are counterexamples, e.g., 3FGL J0212.1+5320 \citep{li2016}. If a redback/black widow has an apparent magnitude of $<15$ mag, it is probably a very nearby pulsar system (i.e. $\leq1$ kpc), and might have been discovered by the pulsar surveys.
\end{itemize}

Ten candidates were selected using the above method. We then further cross-checked the candidates with the SIMBAD\footnote{\url{http://simbad.cds.unistra.fr/simbad/}} database, and found that 6 of them are RR Lyrae variable and 1 of them is a known redback MSP candidate, 3FGL J0954.8$-$3948 \citep{li2018}. We examined visually the CRTS light curves of the rest of the candidates. One of them is an Algol type variable, which has two obvious eclipse dips seen in the light curve. The other two, \first\ and \second, show sinusoidal-like modulations, possibly caused by “pulsar heating” and/or ellipsoidal variation, if they are pulsar systems \citep{heating,bwsize,hui2019,j1048}. We therefore started a multi-wavelength follow-up campaign for the two systems.

\section{4FGL J1910.7$-$5320} \label{j1910}

In the following subsections, we will focus on the results of the multi-wavelength observations of \first.

\subsection{CRTS Surveys Data And VizieR Photometry Viewer} \label{subsec:sed}

We downloaded and re-analyzed the optical data of \first\ from the CRTS \citep{crts} to confirm, and perhaps improve, the period from the CRTS catalog. The phased light curves were also investigated to see if they are consistent with that of a pulsar binary. In addition, the spectral energy distribution (SED) in the optical band obtained from the VizieR Photometry viewer\footnote{\url{http://vizier.unistra.fr/vizier/sed/}} was used to estimate the color temperature of the source as well as the distance to the system.

The CRTS observations were taken from 2005 Aug 01 to 2013 Jun 21, and the variable source is located at R.A.(J2000) = $19^{\rm h}10^{\rm m}49\fs12$, Decl.(J2000) = $-53\arcdeg20\arcmin57\farcs1$. We fit the CRTS data with a sinusoidal function and found that the period is 0.3484776(10) days (roughly 8.36 hours) with a mean magnitude of $19.08\pm0.01$ mag and an amplitude of $0.54\pm0.02$ mag. We note that the period in the CRTS catalog is 0.697 days (twice our best-fit period; \citealt{2017cata}), and at this period, the folded light curve shows a double-peaks feature. Given that the 0.697-day period is inconsistent with the SOAR observations (see \S\ref{sec:soar}), we conclude that the 8.36-hour period is the real one. More detailed investigations will be discussed in the next subsection.

\begin{figure}
\gridline{\fig{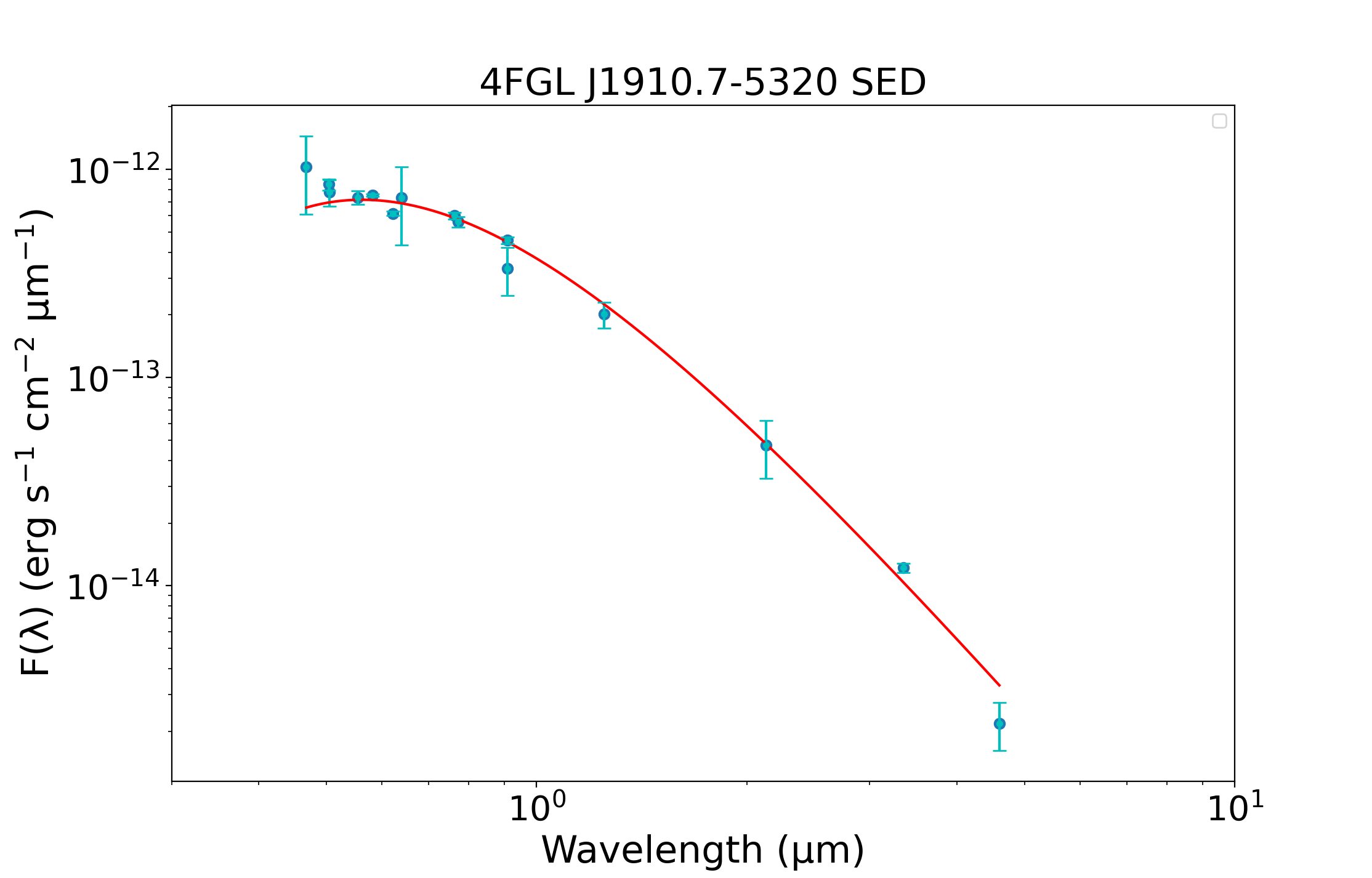}{0.5\textwidth}{}
          }
\vspace{-0.5cm}
\caption{Optical spectral energy distribution (SED) of 4FGL J1910.7$-$5320 with extinction correction. The red line is the best-fit blackbody model.
\label{fig_1}}
\end{figure}

For the SED, we included data from \textit{Gaia} \citep{gaia,gaia2,gaia3}, \textit{GALEX} \citep{galex}, POSS \citep{poss}, VISTA \citep{vista} and \textit{WISE} \citep{wise}, and corrected for the absorption using the extinction function from \citet{extin} with the total absorption in magnitudes of $A_v$ = 0.1857 mag and the ratio of total to selective absorption of $R_v$ = 3.1 \citep{ext}. Then we used a blackbody radiation model to fit the SED data (Figure \ref{fig_1}). The best-fit color temperature is $T=5154\pm164$ K, and the inferred distance is $D = 5.8^{+0.5}_{-0.4}$ ($R/R_\sun$) kpc where $R$ is the companion radius.

Assuming that the companion nearly fills up the Roche-lobe, we used the approximate formula from \cite{egg1983}, which is,
\begin{equation}
 R = a\times\frac{0.49q^{2/3}}{0.6q^{2/3}+\ln(1+q^{1/3})}
\end{equation}
where $q = M_{C}/M_{MSP}$ ($M_{C}$: Mass of companion; $M_{MSP}$: Mass of MSP) and $a$ is the distance between the binary members, to calculate the Roche-lobe radius. By assuming the masses of the MSP and the companion are 1.4 and 0.4 $M_\sun$, respectively, and using Kepler's Third Law to calculate $a$, we find $R \lesssim 0.7 R_\sun$, and hence, $D \lesssim 4.1$ kpc. If we assumed the companion size is similar to that of a black widow (i.e., $M_{C} = 0.03 M_\sun$), then $R \lesssim 0.3R_\sun$ and $D \lesssim 1.8$ kpc.

Nevertheless, the SED data employed were not obtained simultaneously while the optical source is strongly variable. To check whether the effect is huge, we tried to find the blackbody model parameters using only the VISTA J- and K-band data \citep{vista}, which were taken nearly simultaneously. Similar to our original results, the parameters are $T = 4550$ K, $D = 5.2$ ($R/R_\sun$) kpc. The distance is $D \lesssim 3.6$ kpc by assuming the system is redback-like, i.e., $R \lesssim 0.7 R_\sun$.

We also searched the \textit{Gaia} Catalog DR3 \citep{gaia3,gaiaimprove} for further distance information, but the parallax is not well constrained \citep{gaiaest}.

\subsection{SOAR Spectroscopy}
\label{sec:soar}
We obtained optical spectroscopy of the candidate counterpart to \first\ using the Goodman Spectrograph \citep{goodman} on the SOAR telescope over six nights from 2022 Apr 10 to 2022 Jun 10, typically taking multiple spectra per night. For all spectra we used a 1$\farcs$2 slit and a 400\,l\,mm$^{-1}$ grating covering a wavelength range of $\sim 3950$--7850 \AA, giving a resolution of about 7.3 \AA\ for full-width at half-maximum. The spectra were reduced and optimally extracted using standard routines in IRAF \citep{iraf}. We obtained 20 total usable spectra in this setup (Table \ref{tab:tab1}).

\begin{figure}
\epsscale{1.15}
\hspace{-0.5cm}
\plotone{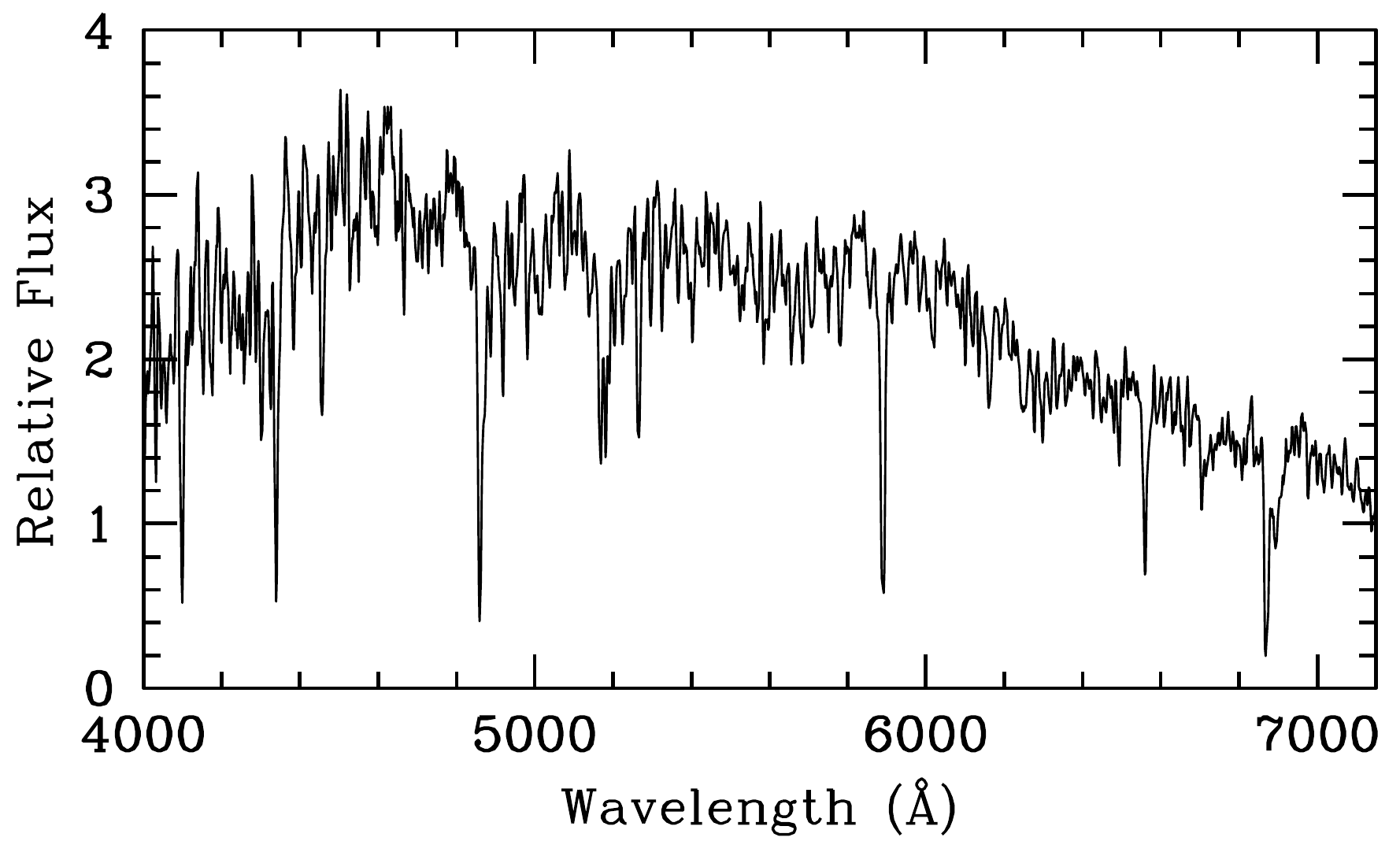}
\caption{A sample spectrum of the optical counterpart to \first\ from 2022 Apr 11. A relative flux calibration has been applied, and the spectrum smoothed with a 3-pixel boxcar for display. Prominent metal and Balmer absorption lines are apparent, as described in \S\ref{sec:soar}. 
\label{fig_2}}
\end{figure}

The spectra generally appear consistent with a late G/early K type star (Figure \ref{fig_2}). The most prominent absorption lines are Mg$b$ and Na D, along with several Fe lines. H$\alpha$ and H$\beta$ are present in absorption in some of the SOAR spectra, while in others H$\alpha$ is weak or absent. There are no clear emission lines in any spectra.

We derived barycentric radial velocities (RVs) through cross-correlation with a high signal-to-noise template spectrum in the region of Mg$b$. We fit a circular Keplerian model to the velocities. As the fitting spectroscopic period is consistent with the photometric period, we fix it to the latter as the time span of the photometry is much longer than that of the spectroscopy. 

The model parameters of the fitting model are $K_{2,obs}=219\pm14$ km\,s$^{-1}$, $\gamma=-17\pm12$ km\,s$^{-1}$, and $T_0=$ BJD 2459700.8091(41), where $K_{2,obs}$ is semi-amplitude, $\gamma$ is systemic velocity, and $T_0$ is the ascending node of the pulsar in Barycentric Julian Date (BJD). This fit has a $\chi^2/$\,degree of freedom (d.o.f.) of 38/17 and a root-mean-square (r.m.s.) of 30.9\,km\,s$^{-1}$, suggesting an imperfect fit. Two of the most negative velocity measurements seem to be unexpected outliers and could have underestimated uncertainties; if these points were excluded, the quality of the fit would be substantially improved. But we have no specific justification for such a change, so we retain the full dataset, and acknowledge this is a preliminary characterization that could be improved with more data in the future.

\begin{figure}
\gridline{\fig{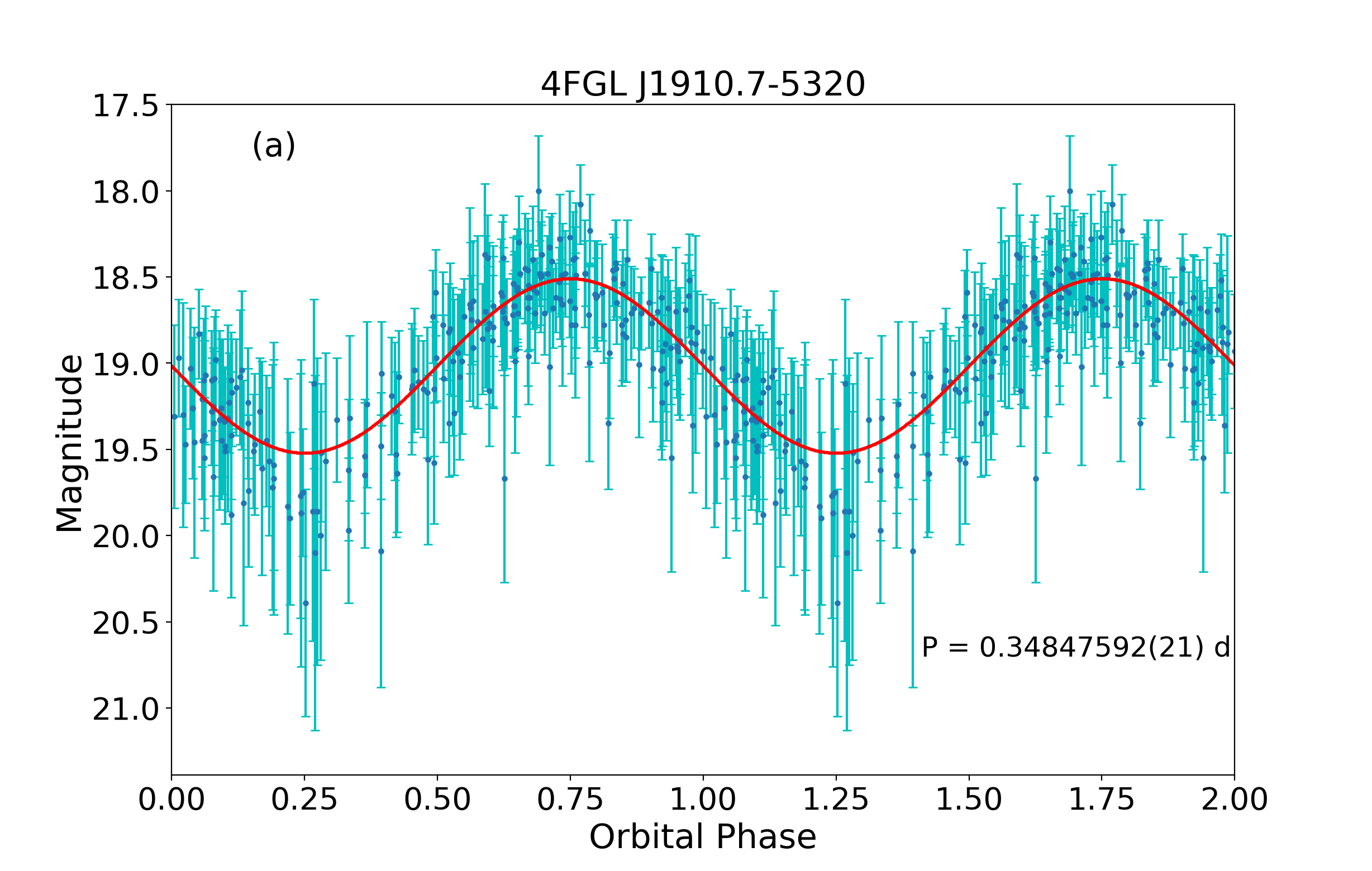}{0.5\textwidth}{}
          }
          \vspace{-0.5cm}
\gridline{\fig{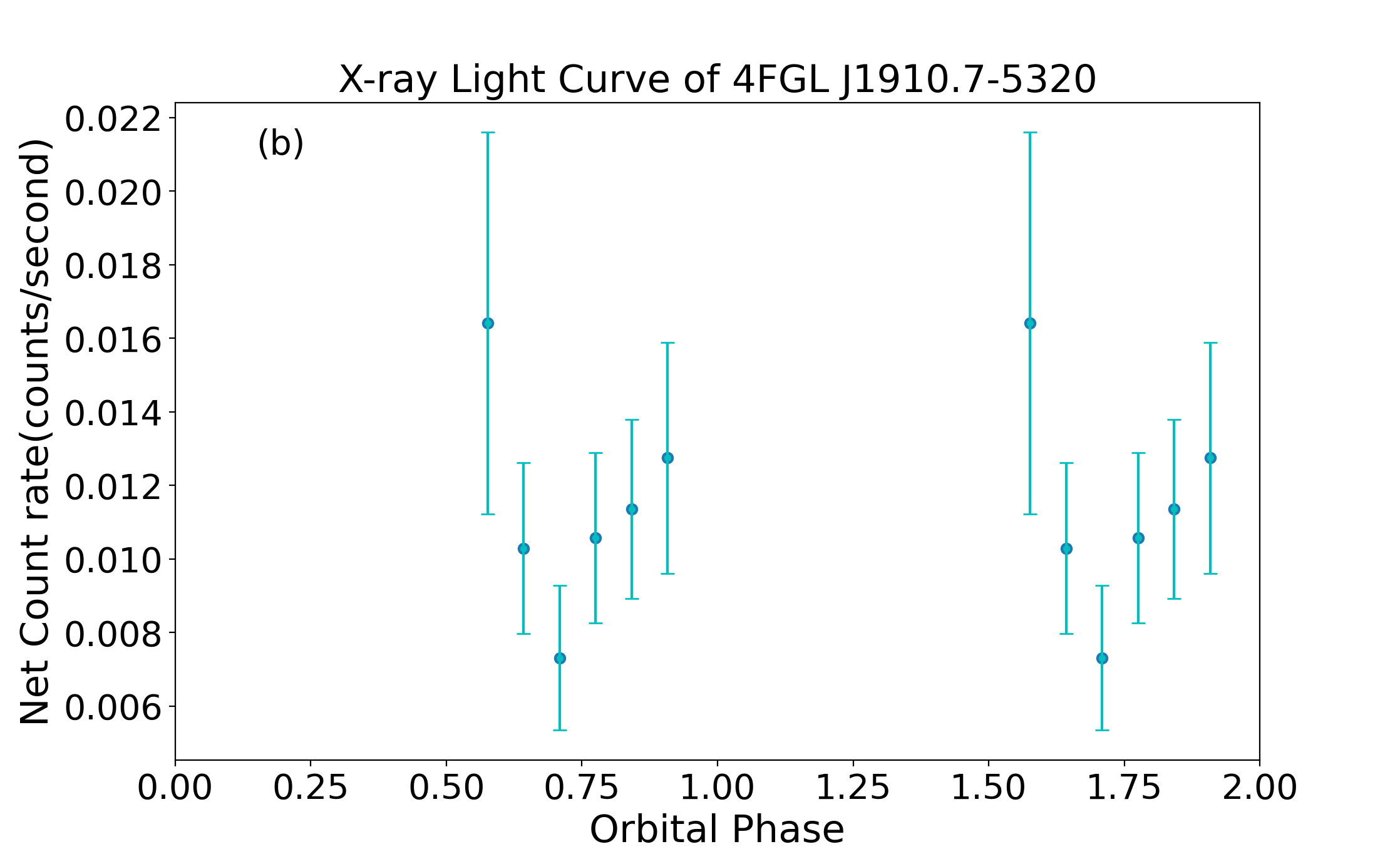}{0.5\textwidth}{}
          }                
\vspace{-0.2cm}
\caption{(a) The CRTS folded light curve and (b) the X-ray folded light curve observed by \textit{Chandra}. The red line in (a) is the best-fit with a sinusoidal  function. Both light curves are folded on the orbital period of P = 0.34847592 days with phase zero at BJD 2453584.0121 which is the ascending node of the pulsar.
\label{fig_3}}
\end{figure}

\begin{figure}
\epsscale{1.2}
\plotone{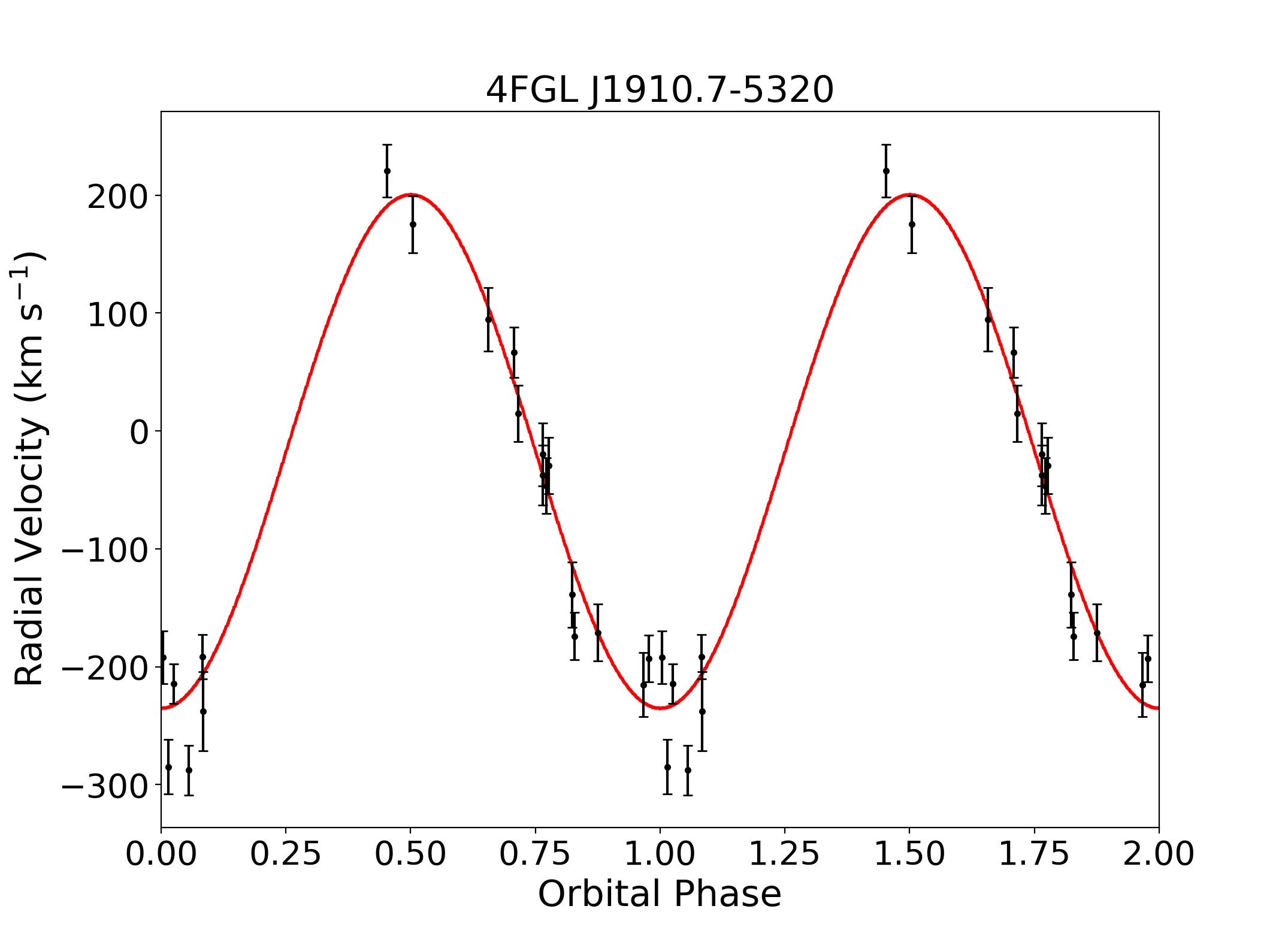}
\caption{Radial velocity (RV) curve of \first. We folded it on the orbital period of P = 0.34847592 days with the phase zero at BJD 2453584.0121.
\label{fig_4}}
\end{figure}

We re-fit both the CRTS and SOAR data assuming a pulsar heating scenario (i.e., the CRTS light curve leads the SOAR RV curve by $\frac{\pi}{2}$). The best-fit parameters are an orbital period of 0.34847592(21) days, a mean CRTS magnitude of $19.015\pm0.018$ mag, a CRTS amplitude of $0.506\pm0.029$ mag, $K_{2,obs} = 218\pm8$ km\,s$^{-1}$, $\gamma = -17\pm6$ km\,s$^{-1}$ and the phase zero at BJD 2453584.0121(31). We use the best-fit parameters of $P = 0.34847592$ days and phase zero at BJD 2453584.0121 to fold both the CRTS light and the RV curves, which are plotted in Figure \ref{fig_3} and \ref{fig_4}, respectively.

It is clear that we have incomplete phase coverage of the source in Figure \ref{fig_3}, missing data around $\phi=0.25$ when the secondary is the faintest. This is not solely chance, but reflects the observational biases induced by the observational window available. Nonetheless, the main orbital parameters are relatively well-constrained.

Because of irradiation, $K_{2,obs}$ is not necessarily the same as the center of mass $K_2$, though no extreme changes are observed in the optical spectra at different phases. The observed mass function implied by the optical spectroscopy is $f = 0.37\pm0.02$ $M_{\odot}$, which is a typical value for a spider binary. Even accounting for the uncertainty in the true $K_2$ value, this mass function (which approximately represents the minimum mass of the primary) implies that an edge-on binary inclination for a neutron star is ruled out, and instead an intermediate inclination is more likely.

\begin{table}
  \begin{center}
  \caption{\label{tab:tab1}}
      \vspace{-0.4cm}
    \begin{tabular}{ c c c } 
      \multicolumn{3}{c}{Radial Velocities of \first\ from SOAR}\\
      \hline
      \hline
      BJD & Radial Vel. & Unc.\\
      (days) & (km s$^{-1}$) & (km s$^{-1}$) \\
      \hline
      2459679.82228114 & $-46.4$ & 23.6\\
      2459679.84012259 & $-138.8$ & 27.6\\
      2459679.85806185 & $-171.1$ & 24.2\\
      2459680.82744858 & $94.5$  &  27.0\\
      2459680.84495725 & $66.9 $ &  31.4\\
      2459680.86457697 & $-19.9 $ & 26.5\\
      2459700.80192541 & $-193.2$& 19.9\\
      2459700.81948426 & $-214.2$ & 16.7\\
      2459700.83923644 & $-191.6$ & 18.9\\
      2459722.66462979 & $15.0$   & 23.9\\
      2459722.68212232 & $-37.5$  & 25.2\\
      2459724.84289214 & $-215.2$ & 27.0\\
      2459724.86042189 & $-284.8$ & 23.3\\
      2459724.88435799 & $-237.7$ & 33.6\\
      2459740.69388734 & $220.7$  & 22.6\\
      2459740.71174963 & $175.3$  & 24.2\\
      2459740.80735656 & $-29.4$  & 23.8\\
      2459740.82483399 & $-174.2$ & 20.1\\
      2459740.88603764 & $-192.1$ & 22.4\\
      2459740.90351553 & $-287.8$ & 21.1\\
      \hline
    \end{tabular}
  
  \end{center}
\end{table}

\subsection{\textit{Chandra} X-ray Analysis} \label{sec:xray}

\first\ was observed with \textit{Chandra} \citep{chandra} for 12.6 ksec on 2019 Nov 20, and we used the \textit{Chandra} data to check whether the CRTS source has an X-ray counterpart. If so, the X-ray data can constrain the X-ray spectral shape and X-ray variability for the putative X-ray counterpart to \first.

We used CIAO (version 4.13; \citealt{ciao}) to extract the source and background spectra from the \textit{Chandra} data. We made an auxiliary response file (ARF; for both source and background) which is an effective area calibration file where we also applied an energy-dependent point-source aperture correction. We generated a response matrix file (RMF) to map between the properties of the incoming photons and the electronic signals obtained from the detector. After performing a spectral binning with at least 20 counts per bin, we used \texttt{XSPEC}\footnote{\url{https://heasarc.gsfc.nasa.gov/xanadu/xspec}} (version 12.12.0; \citealt{xspec}) from HEASARC to measure the hydrogen column density (which will be a fixed parameter in our spectral model) and the photon index of the X-ray source assuming an absorbed power-law model. We also used the \texttt{dmextract} task to generate the light curve with a 2$\arcsec$ radius circular region and 2000\,s bin time. Barycentric corrections were performed using \texttt{axbary}. In Figure \ref{fig_3}, we folded the X-ray light curve using the optical period of $P = 0.34847592$ days.

\begin{figure}
\epsscale{1.15}
\hspace{-0.8cm}
\plotone{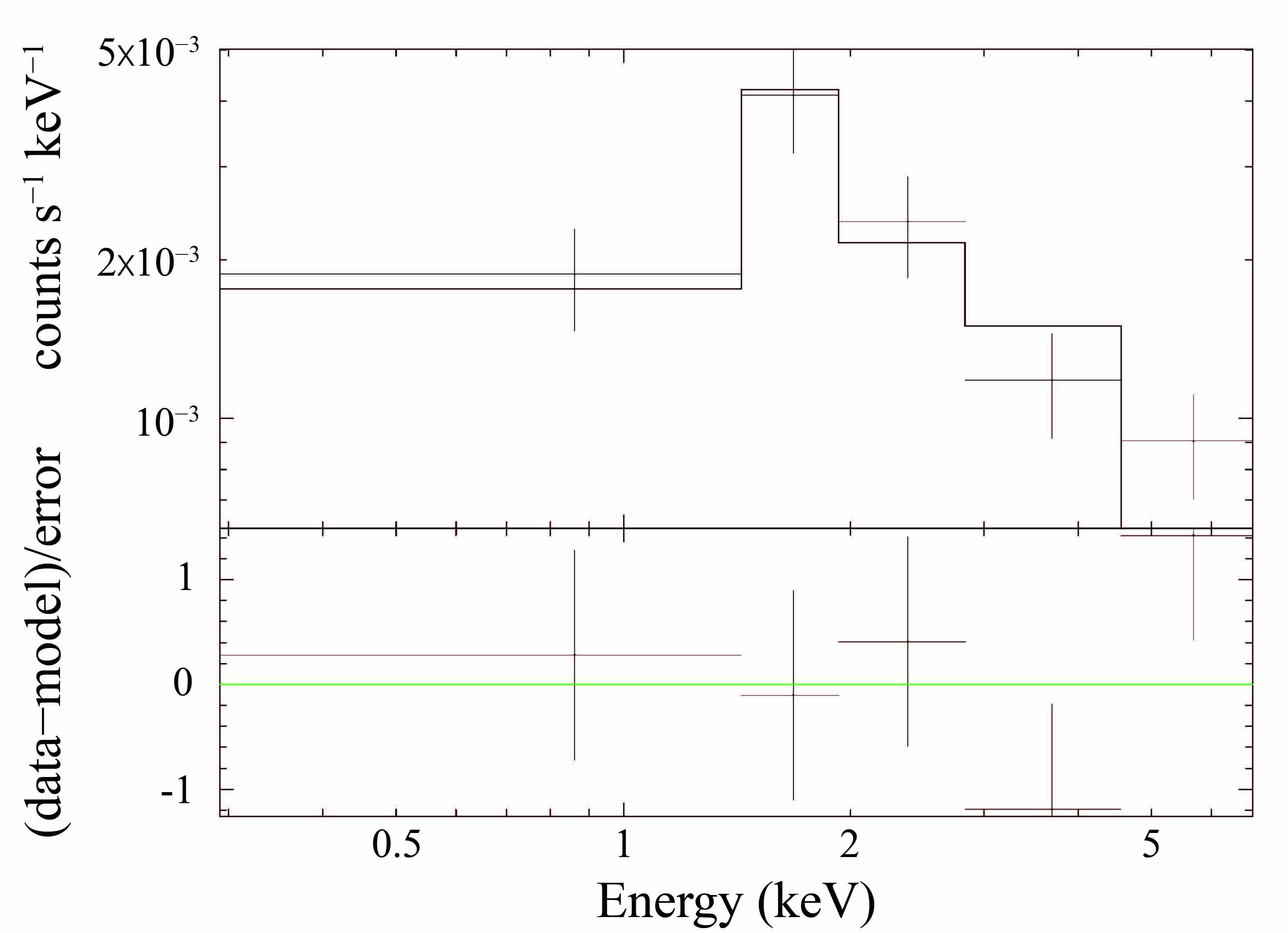}
\caption{The X-ray spectrum in the 0.3-7 keV energy band of \first. It was obtained by the best-fit power-law model. \label{fig_5}}
\end{figure}

A significant X-ray counterpart was detected at the optical position of the variable CRTS source. Its location is at R.A.(J2000) = $19^{\rm h}10^{\rm m}49\fs10$ and Decl.(J2000) = $-53\degr20\arcmin57\farcs2$ with a 90\% uncertainty of 0$\farcs$8\,. This is only 0$\farcs$17 from the CRTS variable described in \S\ref{subsec:sed}, strongly suggesting that they are the same source. There are 106 source counts in a 2$\arcsec$ radius aperture and 164 counts in a nearby source-free circular background region with a radius of 10$\arcsec$. We fit the X-ray spectrum with an absorbed power-law model with the Galactic hydrogen column density of $N_H = 5.22\times10^{20}$ \nh\  (fixed; \citealt{ftools,hi4pi})\footnote{\url{http://heasarc.gsfc.nasa.gov/ftools}}. The best-fit parameters to the X-ray spectrum (Figure \ref{fig_5}) are a photon index of $\Gamma = 1.0\pm0.4$ and an energy flux of $F_{0.3-7keV} = (1.7\pm0.2)\times10^{-13}$ \flux\ ($\chi^2/d.o.f.= 3.7/3$). The 0.3--7 keV X-ray luminosity is $L_x \lesssim (3.4\pm0.4)\times10^{32}$ \lumi\ by assuming $D \lesssim 4.1$ kpc.

We also fit the spectrum with a blackbody model, and the result is $F_{0.3-7keV} = (1.26\pm0.15)\times10^{-13}$ \flux\ and a temperature of $kT = 0.9\pm0.1 $ keV ($\chi^2/d.o.f.= 13.2/3$). The absorbed power-law model is statistically preferred by comparing the $\chi^2$ values.

\begin{figure}
\epsscale{1.2}
\plotone{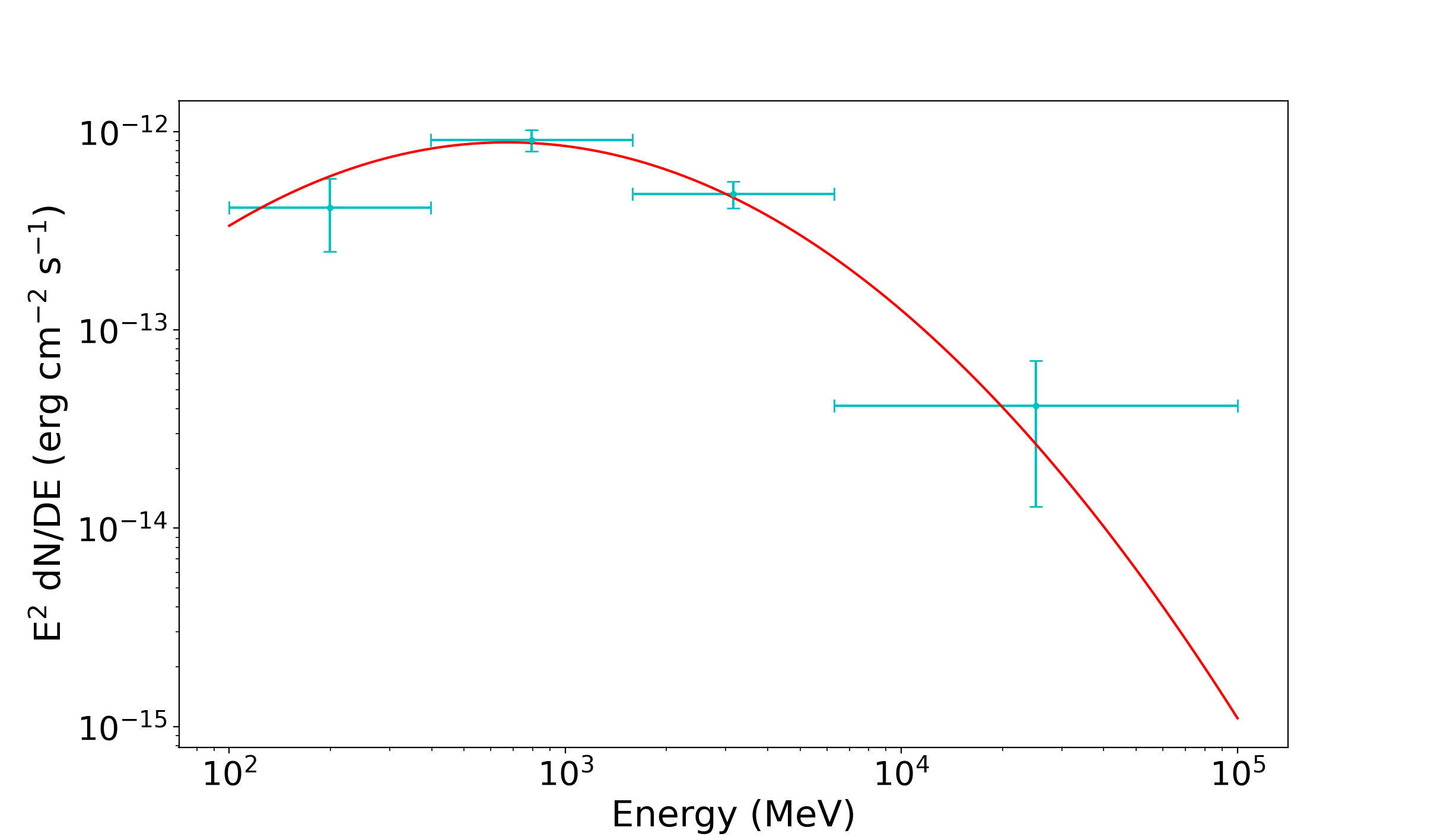}
\caption{The $\gamma$-ray spectrum (\textit{Fermi}-LAT; 0.1--100 GeV) of \first\ with the best-fit LogParabola spectral model indicated by the red line.
\label{fig_6}}
\end{figure}

\begin{figure*}[ht!]
\epsscale{1}
\plotone{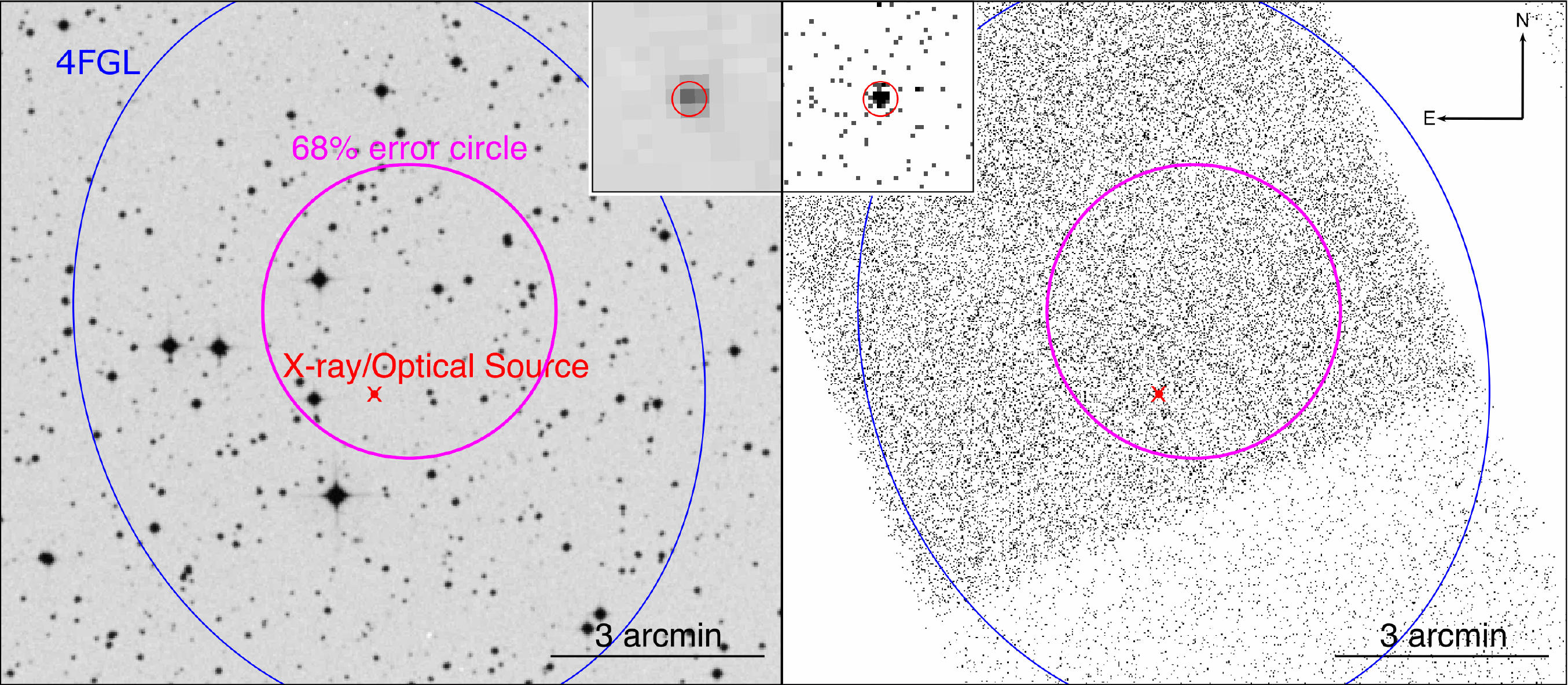}
\caption{The Digitized Sky Survey (DSS) image (left) and the \textit{Chandra} X-ray image (right) of 4FGL J1910.7$-$5320. The blue ellipses show the 95\% error ellipse of 4FGL catalog. The magenta circle is the updated LAT error circle at a 68\% confidence level. The small red cross is the X-ray/Optical position determined by \textit{Chandra} (\S\ref{sec:xray}). The two upper middle inset boxes are the zoomed-in view of the optical and X-ray counterparts.
\label{fig_7}}
\end{figure*}

\begin{figure*}[ht!]
\epsscale{1}
\plotone{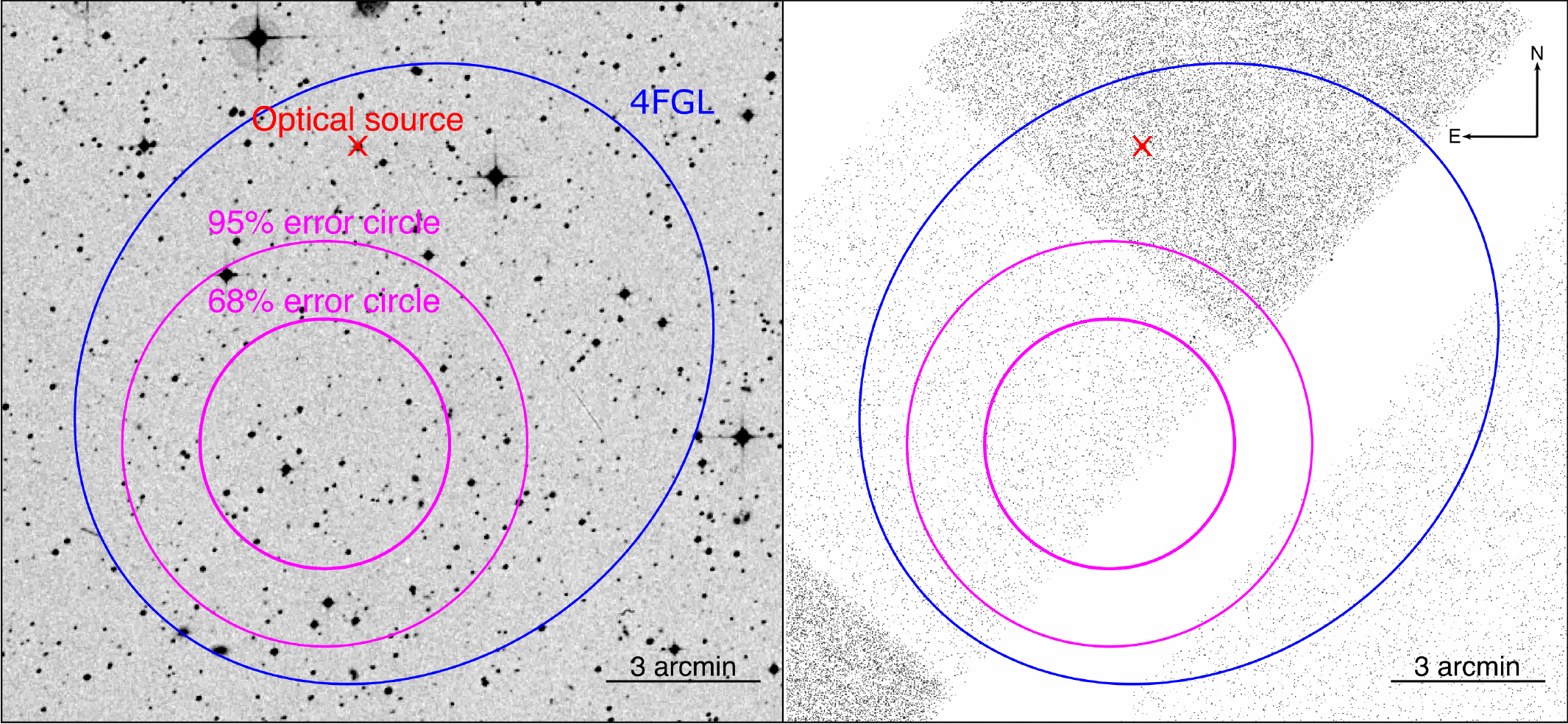}
\caption{The DSS image (left) and the \textit{Chandra} X-ray image (right) of 4FGL J2029.5$-$4237. The blue ellipses show the 95\% error ellipse of the 4FGL catalog. The magenta concentric circles present the updated LAT error circles at a 68\% (inside) and a 95\% (outside) confidence level. The small red cross is the position determined by the CRTS survey.
\label{fig_8}}
\end{figure*}

\subsection{\textit{Fermi}-LAT Gamma-ray Analysis} \label{sec:gammaray}

Here, we used Fermitools (version v11r5p3; \citealt{fermitools}) from the \textit{Fermi} Science Support Center (FSSC)\footnote{\url{https://fermi.gsfc.nasa.gov/ssc}} with the 4FGL-DR3 \citep{fermi4-3} and Pass8 data (P8R3) to refine the $\gamma$-ray position and the $\gamma$-ray spectral properties of \first.

We downloaded the LAT event files and spacecraft data from FSSC. The P8R3 data downloaded starts from 2008 Aug 4 to 2021 Nov 9 with energies in 0.1--300 GeV. We chose the SOURCE class events (FRONT and BACK) with a zenith angle smaller than 90$^{\circ}$. The center of the $14^{\circ} \times 14^{\circ}$ region of interest (ROI) is at ($\alpha$, $\delta$) = ($287^{\circ}.705, -53^{\circ}.349$), the 4FGL-DR3 position of \first. We used the 4FGL-DR3 cataloged sources located within 10$^{\circ}$ from the target to establish the spatial and spectral model of the $\gamma$-ray emission. The model includes the latest Galactic interstellar (gll$\_$iem$\_$v07.fits) and isotropic (iso$\_$P8R3$\_$SOURCE$\_$V3$\_$v1.txt) diffuse components. We employed a LogParabola model for \first\ as suggested in the 4FGL-DR3, which is
\begin{equation}
\frac{dN}{dE} = N_0(\frac{E}{E_b})^{- ( \alpha + \beta \log(\frac{E}{E_b}))},
\end{equation}
where $\alpha$ characterizes the photon index and $\beta$ defines the degree of curvature for the LogParabola model. There is a total of 33 free parameters from the source in the emission model by allowing the background diffuse components and the sources inside a 5$^{\circ}$ radius circle from \first\ to vary. We performed a binned likelihood analysis with 37 logarithmically uniform energy bins, which gives test statistic (TS) value, the significance of a certain source, of 146 ($\sim11.7\sigma$ detection significance with 3 extra parameters), a 0.1--100 GeV energy flux of $F_{0.1-100GeV} = (2.6\pm0.4)\times10^{-12}$ \flux, $\alpha = 2.2\pm0.2$, and $\beta = 0.3\pm0.2$ (Figure \ref{fig_6}). The 0.1--100 GeV $\gamma$-ray luminosity is $L_\gamma \lesssim (5.3\pm0.8) \times 10^{33}$ \lumi\ by assuming $D \lesssim 4.1$ kpc. Using \texttt{gtfindsrc}, we refined the 68\% error circle of \first\ to a circular region with 2$\farcm$1 radius centred at ($\alpha$, $\delta$) = (287$^{\circ}$.691, $-$53$^{\circ}$.330) that includes the CRTS optical source (Figure \ref{fig_7}).

\section{4FGL J2029.5$-$4237} \label{j2029}

The possible optical counterpart to \second\ is located at R.A.(J2000) = $20^{\rm h}29^{\rm m}34\fs21$ and Decl.(J2000) = $-42\degr33\arcmin17\farcs6$, inside the 95\% error region of the $\gamma$-ray source. Following a similar procedure to the analysis of 4FGL J1910.7–5320, the best-fit period of the optical variable is 0.178235614(93) days with a mean magnitude of 14.264$\pm$0.002 mag and an amplitude of 0.183$\pm$0.003 mag. We find no X-ray counterpart in a 11.7 ks \textit{Chandra} observation on 2019 Sep 01, leading to a 3-$\sigma$ flux upper limit of $F_{0.3-7keV} = 2.4\times10^{-14}$ \flux, where we assumed the photon index to be 2 and the Galactic column density of $N_H$ = $3.7\times10^{20}$ \nh\ (fixed; \citealt{ftools,hi4pi})\footnote{\url{http://heasarc.gsfc.nasa.gov/ftools}}. The corresponding 3-$\sigma$ luminosity upper limit is $L_{0.3-7keV} = (3.8\pm0.2)\times10^{30}$ \lumi\ by assuming the \textit{Gaia} DR3 \citep{gaia3} distance of D = 1.14$\pm$0.03 kpc. 

In the \textit{Fermi}-LAT analysis, the P8R3 data used spans from 2008 Aug 4 to 2020 Mar 25 for \second. Using a LogParabola model suggested in 4FGL-DR3, the best-fit TS value is 119 (equivalent to a $>10\sigma$ detection significance), with a 0.1--100 GeV energy flux of $F_{0.1-100GeV} = (2.0\pm0.2)\times10^{-12}$ \flux, $\alpha = 2.4\pm0.2$, and $\beta = 9.998$ (fixed in 4FGL-DR3, likely because it reaches the maximum limit of the parameter space). The updated 68$\%$ error circle of \second\ is located at ($\alpha$, $\delta$) = (307$^{\circ}$.407, $-$42$^{\circ}$.651) with 2$\farcm$4 radius. However, the CRTS optical source is outside this improved 95$\%$ error circle (shown in Figure \ref{fig_8}), strongly suggesting that the CRTS source is unrelated to \second.

\section{Discussion}

We find a candidate optical/X-ray counterpart to \first, and the observational results strongly suggest that it is a redback MSP binary. We summarize the properties of this redback candidate below:

\begin{itemize}
 \item An updated $\gamma$-ray 68\% error circle with 2$\farcm$1 radius is presented in Figure \ref{fig_7} and the optical source is still inside the error circle.
 \item A compact orbit of $P = 8.36$ hours and $K_{2,obs} = 218\pm8$\,km\,s$^{-1}$, which is consistent with the orbital properties of many redbacks \citep{hui2019,jay}.
 \item A single peak orbital light curve of \first\ (Figure \ref{fig_3}) implies that the optical emission probably is pulsar heating-dominated.
 \item An X-ray counterpart to the optical source is also found in \textit{Chandra}. A hard photon index of $\Gamma=1.0\pm0.4$ was shown in X-rays and the emission exhibits tentative evidence for periodic modulation.
\end{itemize}

\subsection{X-ray Orbital Modulation Of 4FGL J1910.7$-$5320}

As we mentioned in \S~\ref{sec:xray} and Figure~\ref{fig_3}, the X-ray and optical sources probably are the counterparts to \first. This association can be further confirmed, if the X-ray and optical phased light curves have any relation.

In order to explore the X-ray modulation, we folded the \textit{Chandra} X-ray light curve on the orbital period P = 0.34847592 days (Figure~\ref{fig_3}). Although the observation does not cover a complete orbit, the folded light curve shows a possible dip at phase 0.75 (Figure~\ref{fig_3}). If the X-ray orbital modulation is real, it could be due to the Doppler boosting effect in the intrabinary shock \citep{li2014,takata2014,kong2017}. At phase 0.75 (Figure~\ref{fig_3}), the stellar companion is behind the presumed pulsar as seen from Earth. If the momentum flux of the pulsar wind is stronger than that of the stellar wind from the companion, the intrabinary shock can wrap the companion. This means the shocked wind does not point towards Earth, and the X-ray emission therefore decreases, consistent with the \textit{Chandra} light curve of \first. Furthermore, the intrabinary shock could produce synchrotron X-ray emission that can be described by a power-law spectral model. This naturally explains the observed \textit{Chandra} X-ray emission, which appears non-thermal with $\Gamma = 1.0\pm0.4$. Unfortunately, the current X-ray and optical datasets do not allow a detailed investigation. Deeper and longer observations in X-rays and the optical band (with color information) are required in the future.

\subsection{4FGL J1910.7$-$5320 As A Redback Or Black Widow?}

From Figure \ref{fig_3}, the optical phased light curve of \first\ shows an obvious one-peak signature. Since the pulsar heating effect is usually more prominent for black widows than for redbacks, this light curve feature is more consistent with that of a black widow. The optical peak and X-ray peak are shifted by half an orbit, which is similar to the original black widow PSR B1957+20 \citep{bw}. However, there are also several black widow examples that show zero phase shift between the optical and X-ray peaks, such as PSR J1124$-$3653, PSR J1653$-$0158 and PSR J2256$-$1024 \citep{gen,1653}. Therefore, the evidence is not particularly strong to say that \first\ is a black widow candidate.

In \S\ref{subsec:sed}, we assumed the size of the companion in the case of redback  (i.e., $R \lesssim 0.7 R_\sun$) and black widow (i.e., $R \lesssim 0.3R_\sun$) based on a Roche-lobe description, and estimated the distance to be $ D \le 4.1$ kpc and $ 1.8$ kpc, respectively. If the system is a black widow MSP, it will be relatively close to us. In fact, the system will be much closer if it is a black widow, because the companion size of a black widow is generally smaller than the Roche-lobe radius \citep{bwsize}. We also performed another independent distance estimation by comparing the magnitude and the distance of \first\ to the original black widow, PSR B1957$+$20, of which the distance is $D =$ 1.5--2.5 kpc \citep{bwdis2,bwdis3,bwdis1} and the minimum optical magnitude in the $R$ band is 24.6 mag \citep{bwlc}. In \first, the minimum best-fit magnitude in the CRTS catalog is 19.62 mag ($\sim$ 20.5 for the faintest data), which is about $\sim$ 5 mag brighter than the minimum magnitude of PSR B1957$+$20. This implies that if \first\ is a black widow MSP similar to PSR B1957+20, the distance to \first\ is $D \sim$ 0.15--0.25 kpc, which is extremely close to us. At this short distance, \textit{Gaia} should be able to measure the parallax, and hence, the distance of 4FGL J1910.7-5320 easily. However, the parallax is not well constrained in the \textit{Gaia} Catalog DR3 \citep{gaia3}. Therefore, interpreting \first\ as a redback MSP system is favored.

\subsection{Strong Irradiation Signature Of 4FGL J1910.7$-$5320?}

We notice that the irradiation signature is relatively strong (the amplitude is over 1 mag) compared to other redbacks. To check this phenomenon systematically, we used the CRTS catalog \citep{crts} to find the modulation amplitude of other known redback systems. We find that there are only two redbacks, PSR J2215$+$5135 \citep{2215} and PSR J2339$-$0533 \citep{2339}, that have a high irradiation signature (amplitude $>$ 1 mag) among the 14 known redbacks \citep{hui2019,jay}. A dedicated statistical study of irradiation signatures of redbacks will be published elsewhere.

\subsection{4FGL J2029.5$-$4237}

We found a variable optical source inside the 95\% error ellipse of \second\ and speculated that it is another MSP binary candidate. In the 11.7 ks \textit{Chandra} observation, we found no significant X-ray source spatially coincident with the optical variable, with a 3-$\sigma$ flux upper limit of $2.4\times10^{-14}$ \flux\ (0.3--7 keV). The corresponding luminosity is $L = 3.8\times10^{30}$ \lumi\ by using the \textit{Gaia} distance of 1.14 kpc \citep{gaia3}, which is considerably lower than in many other MSPs. The optical source is also outside the updated 95\% LAT error circle (Figure \ref{fig_8}). These results strongly suggest that the variable optical source is probably not the counterpart to \second\ and unlikely a redback MSP.

\hspace{1cm}

While we were in the late stages of preparing this paper, TRAPUM discovered the radio pulsations associated with \first\ (i.e., PSR J1910$-$5320; private communication)\footnote{\url{http://www.trapum.org/discoveries/}}, and confirms that \first\ is a redback MSP. They also found that \second\ is an isolated MSP, which is unrelated to the optical source reported in this paper.

\newpage

\begin{acknowledgments}

The \textit{Fermi}-LAT Collaboration acknowledges generous ongoing support from a number of agencies and institutes that have supported both the development and the operation of the LAT as well as scientific data analysis. These include the National Aeronautics and Space Administration and the Department of Energy in the United States, the Commissariat \`{a} l'Energie Atomique and the Centre National de la Recherche Scientifique / Institut National de Physique Nucl\'{e}aire et de Physique des Particules in France, the Agenzia Spaziale Italiana and the Istituto Nazionale di Fisica Nucleare in Italy, the Ministry of Education, Culture, Sports, Science and Technology (MEXT), High Energy Accelerator Research Organization (KEK) and Japan Aerospace Exploration Agency (JAXA) in Japan, and the K. A. Wallenberg Foundation, the Swedish Research Council and the Swedish National Space Board in Sweden. Additional support for science analysis during the operations phase from the following agencies is also gratefully acknowledged: the Istituto Nazionale di Astrofisica in Italy and and the Centre National d'Etudes Spatiales in France. This work performed in part under DOE Contract DE-AC02-76SF00515.

This research has made use of data obtained from the \textit{Chandra} Data Archive and the \textit{Chandra} Source Catalog, and software provided by the \textit{Chandra} X-ray Center (CXC) in the application packages CIAO and Sherpa.

Based on observations obtained at the Southern Astrophysical Research (SOAR) telescope, which is a joint project of the Minist\'{e}rio da Ci\^{e}ncia, Tecnologia e Inova\c{c}\~{o}es (MCTI/LNA) do Brasil, the US National Science Foundation’s NOIRLab, the University of North Carolina at Chapel Hill (UNC), and Michigan State University (MSU).

The CRTS survey is supported by the U.S. National Science Foundation under grants AST- 0909182 and AST-1313422.

KYA and KLL are supported by the National Science and Technology Council of the Republic of China (Taiwan) through grant 111-2636-M-006-024, and KLL is also a Yushan Young Fellow supported by the Ministry of Education of the Republic of China (Taiwan).

JS acknowledges support by NSF grant AST-2205550 and the Packard Foundation. This research was performed while SJS held a NRC Research Associateship award at the Naval Research Laboratory. Work at the Naval Research Laboratory is supported by NASA DPR S-15633-Y.

CYH is supported by the National Research Foundation of Korea through grants 2016R1A5A1013277 and 2022R1F1A1073952.

\end{acknowledgments}

\facilities{\textit{Fermi}, CXO, SOAR}
\software{CIAO (version 4.13; \citealt{ciao}), HEASOFT \citep{heasoft}, FERMITOOLS (version v11r5p3; \citealt{fermitools})}

\bibliographystyle{aasjournal}


\end{document}